\documentclass[english,twocolumn,aps,prl,groupedaddress,showpacs]{revtex4}
\usepackage[T1]{fontenc}
\usepackage[latin1]{inputenc}
\usepackage{subfigure}
\usepackage{graphicx}
\usepackage{amssymb}

\makeatletter

\usepackage[T1]{fontenc}
\usepackage[latin1]{inputenc}
\usepackage{graphicx}
\usepackage{amssymb}

\makeatletter

\usepackage{natbib}
\usepackage{graphicx}
\usepackage{bm}
\usepackage{amsfonts}
\usepackage{amssymb}
\usepackage{amsmath}

\makeatother
\makeatother
\makeatother
\makeatother
\makeatother
\makeatother

\usepackage{babel}
\makeatother

\usepackage{babel}
\makeatother
\begin{document}

\title{Interacting damage models mapped onto Ising and percolation models}

\author{Renaud Toussaint}

\email{Renaud.Toussaint@fys.uio.no}

\homepage{http://folk.uio.no/renaud/}

\affiliation{Department of Physics, University of Oslo, POBox 1043 Blindern, 0316
Oslo, Norway.}

\author{Steven R. Pride}

\email{srpride@lbl.gov}

\affiliation{Lawrence Berkeley National Laboratory, Earth Science Division, 1
Cyclotron Road, MS 90-1116, Berkeley, CA 94720, USA.}

\date{\today{}}

\begin{abstract}
We introduce a class of damage models on regular lattices with isotropic 
interactions between the broken cells of the lattice.   
Quasistatic fiber bundles are an example. The interactions
are assumed to be weak, in the sense that the stress perturbation from a 
broken cell is much smaller than the mean stress in the system. 
 The system starts intact with a  
 surface-energy threshold required to break any cell
sampled from an uncorrelated quenched-disorder distribution.
The evolution of this heterogeneous system is ruled by Griffith's
principle which states that a cell breaks  when the release in potential (elastic)
energy in the system exceeds the surface-energy barrier necessary to  
break the cell. By direct integration over all  possible realizations of the quenched disorder,
we obtain the probability distribution of each damage configuration at any level
of the imposed external deformation. We demonstrate an isomorphism  
between the  distributions so obtained and standard generalized Ising
models, in which the coupling constants and effective temperature in the 
Ising model are functions of 
the nature of the quenched-disorder distribution and  the extent of accumulated damage.
In particular, we show that damage models with global load sharing
 are isomorphic to standard percolation theory,  that damage
models with local load sharing rule are isomorphic to the standard Ising
model, and draw consequences thereof for the universality class and
behavior of the autocorrelation length of the breakdown transitions
corresponding to these models. We also treat damage
models having more  general power-law  interactions, and classify the
breakdown process as a function of the power-law interaction exponent. Last, we also
show that the probability distribution over configurations is a maximum of Shannon's
entropy under some specific constraints related to the energetic balance
of the fracture process, which firmly relates this type of 
quenched-disorder based damage model to standard statistical mechanics.
\end{abstract}



\maketitle

\section{Introduction}

\label{sec:Introduction} 
The physics of breakdown processes that lead, for example,
to stress-induced catastrophic failure of man-made and geological
structures, remains an ongoing subject of research.
Stress-induced  fracture  of a homogeneous material containing
 a geometrically simple single flaw has been studied since
 the work of Griffith \cite{Griffith20} and is now well understood.
However, the breakdown of heterogeneous structures, in which the
local mechanical properties are randomly distributed in space and/or time,
continues to present challenges despite the many advances over the
last fifteen years
\cite{HerrmannR90}.
The difficulty is in characterizing and quantifying the effects
of interaction between the multitude of constituents.

 Most of the knowledge
about these types of systems has been obtained from lattice network simulations.
One of the most well-studied lattice network models is the 80 year old
Fiber Bundle Model
(FBM)  \cite{Daniels45,Daniels89,Coleman58,Peirce26}
that describes the rupture of bundles of parallel fibers. This
model originally considered elastic fibers of identical elastic constant, breaking
when their elongation exceeds individual thresholds distributed according
to a given uncorrelated random distribution.  A global load sharing rule (GLS)
is assumed by which
the load carried by a fiber is uniformly distributed to  the surviving fibers when
it breaks. Analytical results for this model have been obtained
 \cite{Sornette89,Sornette92,Andersen97,HemmerHansen92,PrideT02} that concern
the average load/deformation properties \cite{Sornette89,Sornette92,Andersen97},
the distribution of avalanches \cite{HemmerHansen92},  or the relationship
between such quenched-disorder based models and standard statistical
mechanics  \cite{PrideT02}.

The original FBM  model with global-load sharing has
been generalized to allow for  non-uniform
load sharing rules, considered either as purely local load sharing (LLS), in which
case distribution of avalanches and mechanical properties have been
analytically studied \cite{HansenHemmer94,Kloster97,Zhang96}, or
as power laws of distance from the failed fiber, which have been studied
numerically \cite{Hidalgoetal02,Yewandeetal02}. The distribution of avalanche size
$s$ has been shown to be a power law $s^{-5/2}$ close to macroscopic
breakdown in GLS, and not to follow any power-law in LLS \cite{Kloster97,HansenHemmer94}.


Closely related to these models of fiber bundles in elastic interaction,
many studies have addressed fuse networks 
\cite{Bakke03,BatrouniH98,DeArcangelisH89,DeArcangelisHHR89,DuxburyLB87,Hansen03,HansenHR91b,Zapperi00,Zapperi03},
or networks of isotropic damage with interactions having the range of    
 the elastostatic Green function \cite{BatrouniH02,Herrmann89,Schmittbuhl03}.
Such work has recently helped to understand the origin of a universal geometric
feature of crack surfaces;  namely, their large-scale self-affinity. 
It  is now established that three-dimensional fracture surfaces in disordered
brittle solids are self-affine, with a material-independent roughness
exponent $\zeta\sim0.8$ at large scales 
\cite{Bouchaud97,Schmittbuhl95b,Schmittbuhl93,Cox93,Maloy92,Bouchaud90},
and a cross-over to $\zeta\sim0.5$ at smaller scales. \cite{Daguier96,Daguier97}
\footnote{apart from sandstone where the fracturing process is mostly intergranular,
which displays $\zeta\sim0.5$ at scales up to metric: Y. Meheust, PhD
Thesis, ENS Paris, 2002.}
 In another load geometry, interfacial crack pinning between two
sintered PMMA plates have been reported to give rise to crack fronts
with $\zeta\sim0.6$ \cite{Schmittbuhl97,Delaplace99}.
Recent work \cite{Hansen03,Schmittbuhl03} has explained
 these large-scale roughness exponents using 
 gradient-percolation  theory (\cite{Sapoval85}, relationship to fracture problems first introduced in Refs.\ 
\cite{Zapperi00,Zapperi03}), together with estimates of 
the correlation-length divergence
exponent obtained  using  
 finite-size scaling of lattice models in the approach to failure. 
Scaling laws were produced relating the 
roughness exponents of fracture or damage fronts,  to the 
 critical exponent describing the divergence of an autocorrelation length in 
an appropriate damage model.  
Thus, one important objective of the present work is to analytically   
extract such divergence exponents for a broad class of damage models 
and to show how such exponents depend on the  model characteristics. 

The classification of lattice breakdown processes 
as  critical-point phenomena is still subject to debate
\cite{Moreno01,MorenoGP00,ZapperiRS+97,ZapperiRS+99,ZapperiVS97,Kun00,Andersen97,Sornette98,Caldarelli96,Raisanen98}.
Local load sharing  models are usually understood as breaking
through a process similar to a first-order transition \cite{Moreno01},
while models with long-range elastic interactions or GLS are analyzed either
as a critical-point  transition \cite{Moreno01,MorenoGP00,Sornette98,Andersen97},
or as a spinodal nucleation process \cite{ZapperiRS+97,RundleK89}.
The issue depends on whether
the accumulated damage has a correlation length that diverges as a power
law of the average deformation in the approach to macroscopic failure.
Numerical evidence in the literature suggests
that the nature of the  correlation length in the approach to failure
depends on the specific model being analyzed, the damage interaction range, and
 the type of quenched-disorder distribution considered.
A major difficulty of such attempts to classify the breakdown process is that
no  analytic form for the
distribution of the damage configurations at a given external load
is available (at least that has  a firm basis).
 Numerical simulations of the scaling behavior of avalanches
are limited in size due to
computational restrictions which makes critical-point analysis difficult.

It is proven here that quasi-static interacting damage models,
that possess local breaking thresholds randomly quenched
{\em ab initio}, can be mapped onto percolation, Ising or generalized
Ising models with non-zero coupling constants for distant cells, depending
on the range of the assumed load sharing rule  and on the specific
probability distribution of
 the breaking thresholds. The stress-induced emergent-damage distribution
 is proven here to be a Boltzmannian with a
 temperature (probabilistic energy scale) that is an
 explicit  analytical function of the applied
external load.   Having an analytical form for the
damage-state distribution function allows  a rigorous classification of
 the breakdown process into first-order  or  critical-point phase transitions
belonging to various universality classes.

We have recently proposed a statistical theory for the localization
of oriented fractures that emerge and elastically interact when the system has
a shear stress applied to it
 \cite{ToussaintPr02a,ToussaintPr02b,ToussaintPr02c}.
The distribution of the emergent crack states was obtained using the postulate
that the fracture arrival
would maximize Shannon's entropy under constraints representing the
energetic balance of the process. In the present work, we will not
make this postulate but will instead prove its validity by
direct integration over the  damage evolution.
The present work  also presents more general ranges of interaction.  However, unlike
our previous work, the present analysis is for a purely scalar
description of damage interaction.  The interaction of real fractures in an anisotropic load, where microfractures present high aspect ratios, 
 necessarily requires a tensorial elastic description as our earlier work
\cite{ToussaintPr02a,ToussaintPr02b,ToussaintPr02c} provided.

Other analytical treatments
of  fracture processes in disordered brittle
materials can be found in the literature \cite{RundleK89,BlumbergSelingerWGBS91}.
These approaches  directly postulate both the appropriateness of classical
Boltzmann distributions and the form of a  free energy
containing  non-local long-range interactions.  Such approaches have led to the
analysis of fracture processes as a spinodal nucleation.
However, the mechanical origin of such a free energy and its relation
to the underlying presence of fractures and heterogeneity in the system,
as well as the physical meaning
of the temperature parameter, remain unclear in these models.
Fractures usually form irreversibly, which makes such direct applications
of standard equilibrium statistical mechanics questionable. Fracture
theories as a thermally activated process \cite{Griffith20,Pomeau92}
have long been proposed.  However,  the
 pertinence  of thermal activation in highly heterogeneous and stiff systems
is doubtful, because the
thermal energy is negligeable compared to the gaps between the
strain energies of the possible states of the system.

Thermally induced fracture models \cite{Pomeau92}, generalized with quenched disorder in the
rupture thresholds \cite{Roux00,Politi02,Santucci03,Ciliberto01}, 
 have recently been confronted
with experiments.  Microacoustic
emissions in heterogeneous materials prior to failure  were experimentally
observed to have a cumulative
energy that follows a  power law in time
to failure, while  the individual microacoustics events have energies distributed as
a power law
\cite{Guarino02,GuarinoGC98,GarcimartinGBC97}. In models reproducing
these experimental results through time-delayed fiber bundles
with GLS, it has been pointed out that the type of quenched-disorder
distribution assumed for  the individual failure thresholds directly
 influences  the
equivalent temperature in  thermally activated rupture models
\cite{Roux00,Politi02,Santucci03,Ciliberto01},  which
makes this temperature significantly different from the usual  temperature
associated with  molecular motion.

In comparison with these latter works, the models  developed
here are quasistatic and do not include any time-delayed fracture
processes, but their advantage is the consideration of an arbitrary range
of interactions and of arbitrary  distribution functions for the fracture
thresholds (similar to Refs. \cite{Hidalgoetal02,Yewandeetal02},
except that the present work is analytical rather than numerical).
The nature of the energy scale (called temperature)
that enters the  probability distribution will also be clarified.

The organization of the paper is as follows: in section \ref{sec:Definition-of-the},
we introduce (and justify in the appendix) 
the general type of scalar damage models that
are considered.
In section \ref{sec:Obtention-of-the},  the probability
of each damage configuration is obtained by integrating
over all paths that lead to it.
In section \ref{sec:Equivalence-of-the}, we establish  the relationship
between these configurational distributions and standard statistical
mechanics, which allows  the  standard toolbox of statistical mechanics
to be applied to damage models entirely based on quenched disorder.
These analytical developments will then be utilized in section
\ref{sec:Applications}  to establish that fracture processes in such
damage models are isomorphic to percolation  for GLS (\ref{sub:Global-load-model}), or 
to the Ising
model for LLS (\ref{sub:Local-load-model}). This  allows us to isolate
some transition points in these models, and to predict 
 the nature of the correlation length in the approach to the 
 transition. We will also discuss in Section \ref{sub:Power-law-decay}
the case of damage models with arbitrary power-law decay of the interactions,
and show how they are related to generalized Ising models, which can
themself be mapped onto standard ones via renormalization of the coupling
constants. The results are summarized and discussed in a concluding section.

\section{Definition of the damage models considered}

\label{sec:Definition-of-the} 

Our models reside on 
a   regular lattice of dimension $D$  
(e.g., a square lattice in $D=2$).  Each cell of the lattice 
 has a location $x$  within the ensemble 
$\Omega$ of cells making up the lattice, and has a state 
described by a local order parameter $\varphi_x$, where 
$\varphi_x = 0$ if the cell is intact and $\varphi_x = 1$ 
if it is broken.  There is a local stress and strain 
associated with each cell.  The cells  elastically 
interact with each other; however, such interaction must 
be isotropic for the present analysis to apply.   
A configuration of damage is described 
as a damage field $\varphi$ corresponding to the set of $N$ 
 local variables $\varphi\equiv\left\{ \varphi_{x}\right\} _{x\in\Omega}$
where $N=\text{Card}(\Omega)$ is the total number of cells in
the system.

Our systems are  initially uniform by hypothesis; i.e., they have  
a homogeneous damage field $\varphi = 0$ at zero strain and stress 
and each cell starts with the same elastic moduli. 
Strain is progressively applied  through the application 
 of a uniform normal displacement $l$ at the edges of the system.   
A  cell breaks at constant applied $l$  
when the energy required to break it (which is a random quenched threshold 
sampled from a  probability distribution function) 
just equals the reduction in stored elastic energy in the lattice 
due to the break.    

A key requirement of the class of models 
treated here is that the stress perturbation 
emanating from a broken cell must be weak enough 
that a first Born approximation holds.  This means that  
the stress interaction between any two broken cells is allowed for while 
simultaneous interaction between three or more broken cells is not.  
This approximation is valid whenever the stress perturbation due to a 
broken site is much smaller than the mean stress in the system.  

One specific realization of such a ``weak damage'' model is an appropriately 
defined fiber bundle model.  In the model, a set of $N$ elastic fibers are 
stretched between a free rigid plate and an elastic halfspace. The 
rigid plate is displaced by a controlled amount $l$ that puts the fibers 
in tension.  Once fibers begin to break, elastic interactions occur 
through the elastic solid.  As demonstrated in the appendix, 
such interactions will be weak if either: (1)  
the elastic solid is much stiffer than the fiber material; or, (2) the fibers 
are much longer than wide and are sufficiently widely placed. 
In the limit that the elastic halfspace becomes rigid, this model reduces 
to the classic global-load sharing fiber bundle. 

Another realization is a uniform elastic isotropic solid divided into $N$ cells. 
Uniform displacements  $l$ are applied normally to the limiting 
faces of the lattice  in such a way  that the material is in  
a state of uniform dilation. 
The damage that arrives is assumed to change the isotropic moduli without 
creating anisotropy in the process. For example, the damage might 
be modeled as  a spherical cavity that opens at the center of the cell 
thus reducing the elastic moduli of that cell. The weak interaction is guaranteed 
if the change in the cell modulus is small.  This can be considered 
a special case of the oriented damage models we have considered in 
earlier work \cite{ToussaintPr02a,ToussaintPr02b,ToussaintPr02c}. 

As demonstrated in the appendix,  
the energy $E_p$  reversibly stored in such systems when the system is  
in a damage state $\varphi$ at the applied loading level $l$ is 
\begin{eqnarray}
E_{p}[\varphi,l] & = & (C_{0}+C_{1}+C_{2})l^{2}\label{eq:E,pot}\\
C_{0} & = & N\label{eq:comp,0}\\
C_{1} & = & - c \sum_{x\in\Omega}\varphi_{x}\label{eq:comp,1}\\
C_{2} & = & -\varepsilon\sum_{x,y}J_{xy}\varphi_{x}\varphi_{y}\label{eq:comp,2}\end{eqnarray}
where $c$ is a  positive constant in the range $0< c \le 1$ that is 
independent of the damage/deformation state, 
 $\varepsilon$ is a small positive parameter in the range $0<\varepsilon/c \ll 1 $  
that controls the strength of the stress perturbations, 
and $J_{xy}$ are $O(1)$ coupling
constants  that allow for  the load redistribution between
cells at positions $x$ and $y$  when a cell is
broken. Various spatial ranges for $J_{xy}$ are  considered including:  
 (1) global load sharing 
in which case $J_{xy}=0$;  
(2) local load sharing  in which
case $J_{xy}=\alpha>0$ is a constant of order unity for pairs $xy$
of nearest neighbors, $0$ otherwise;   
(3) elastic load sharing in  
which case $J_{xy}\sim(\ell/|x-y|)^{D}$ where 
$\ell$ is    
 the  lattice step size;  and  
(4) fiber-bundle elastic sharing  with  fibers interacting through an elastic plate  
 in which case 
$D=2$ and $J_{xy}\sim \ell/|x-y|$ [see the appendix].
 
The local load sharing case (2) can in principle happen in a fiber bundle stretched between  
plates enduring both elastic and plastic deformations capable of  
screening the stress perturbations caused by  broken fibers to only nearest neighbors, 
which always can carry some load if damaging them only decreases their elastic constant 
(i.e. $c < 1$). We treat case (2) for the sake of generality and 
do not specify the detailed constitutive relations required for it to be realized in practice.

The cost in surface energy to break any of the cells (which represents either 
the energy required to create new surface area within  a cell or to break a fiber) 
is a random variable
fixed {\em ab initio}, with no spatial correlations between the different
cells.  The breaking energy is thus a quenched uncorrelated disorder, described by a probability
distribution function $p(e)$ for  which  $p(e)de$ is the probability that  a cell's surface
energy is in $[e,e+de]$, and having a cumulative distribution $P(e)=\int_{0}^{e}p(z)dz$.
For a given realization of each cell's surface energy $e_{x}$,
there is thus a certain total surface energy 
\begin{equation}
E_{s}[\varphi]=\sum_{x\in\Omega}e_{x}(1-\varphi_{x})\label{eq:E,surf}\end{equation}
associated with  each damage
field $\varphi=\left\{ \varphi_{x}\right\} _{x\in\Omega}$.

Given that $B\subset\Omega$ is a certain subset of locations, the
notation $\varphi^{B}$ refers to the state where this subset is broken
and its complementary intact; i.e.,  $\varphi_{x}^{B}=1$ for every
$x\in B$, and $\varphi_{x}^{I}=0$ for every $x\in I=B^{C}$.

As the external deformation $l$ is increased, damage evolution  
is ruled by Griffith's principle:    
  Given that the system is in a certain
configuration $\varphi^{B}$, at a given deformation $l$, it can
undergo a transition towards a more broken state $\varphi^{B\cup\left\{ x\right\} }$
differing from the previous one by one additional broken cell 
at $x$,  
 if the release in potential energy is equal to
the surface energy cost of the new break;  i.e., if 
\begin{equation}
\Delta E_{p}[\varphi^{B},x,l]  =  e_{x}\label{eq:cond,break}
\end{equation}
where 
\begin{eqnarray}
\Delta E_{p}[\varphi^{B},x,l] & = & E_{p}[\varphi^{B},l]-E_{p}[\varphi^{B\cup
\left\{ x\right\} },l]\nonumber \\
 & = & (c+\varepsilon\sum_{y}J_{xy}\varphi_{y})l^{2}.\label{eq:expr,delta,E,p}
\end{eqnarray}
If $e_{x}>\Delta E_{p}[\varphi^{B},x,l]$ for any surviving cell $x\in I$,
there is no break and the deformation can be further increased while
the system remains in the same state $\varphi^{B}$. If a break happens
in cell $x$, which leads to  the new state $\varphi^{E}$ where $E=B\cup\left\{ x\right\} $ 
while the external deformation $l$ is kept constant,
 there is a possibility
of avalanche at fixed $l$ if there is some $y\in E^{C}$ such that 
\begin{equation}
\Delta E_{p}[\varphi^{E},y,l]\geq e_{y}.\label{eq:cond,aval}
\end{equation}
If there is more than one possible location satisfying Eq.\ (\ref{eq:cond,aval}), 
the one which breaks is determined by maximizing the energy release; 
i.e., its location corresponds to 
\begin{equation}
\Delta E_{p}[\varphi^{E},y,l]-e_{y}=\label{eq:avalanche,location,determined}
 \text{Max}_{z\in E^{C}}\left(\Delta E_{p}[\varphi^{E},z,l]-e_{z}\right).  
\end{equation}
The avalanche test [Eq.\ (\ref{eq:cond,aval})] is then computed
again until the system stabilizes in a given configuration.

Eventually, although we have chosen to base the evolution of our damage
model on minimization of energy, we note that the formal integration
of path probabilities presented in the following sections could similarly
be obtained as well for the case of a rule based on force thresholds,
with at any given external deformation, a force carried per intact
fiber equal to an average one, plus perturbations due to the already
broken fibers. 
However, this approach will not be pursued here.

\section{Probability distribution of damage states}

\label{sec:Obtention-of-the}

The probability of occurence of any damage configuration
$\varphi$ is now determined 
when the system is at a given external deformation $l$ that was  reached
monotonically ($l$ as defined here does not include any elastic unloading). 
We will first summarize the results for the simplest
case,  global load sharing, which was performed in \cite{PrideT02},
and which will serve as a basis for a perturbative treatment to include
the effect of interactions.

\subsection{Global load sharing \label{sub:No-interactions--}}
In this case, the interaction term of
Eq.\ (\ref{eq:comp,2}) is $C_{2}=0$ for any configuration, and from
Eq.\ (\ref{eq:expr,delta,E,p}), $\Delta E_{p}[\varphi,x,l]=c l^{2}$
regardless of the  state $\varphi$ and new break location $x$ considered.

Each of the cells then share the same level of deformation $l$, and
the probability for any one  of them to be broken is simply (from the cumulative 
surface energy distribution)  
\begin{equation}
P_{0}=P(e<c l^{2}) = \int_0^{cl^2} p(z) \,dz.  \label{eq:P,0}\end{equation}
The probability of being in a configuration $\varphi$ with
$n$ cells broken out of $N$ is then 
\begin{equation}
P[\varphi,l]=P_{0}(l)^{n}(1-P_{0}(l))^{N-n}.\label{eq:P,phi,glob,noint}\end{equation}
This corresponds to the behavior of a so-called ``two-state system'' in which 
each of the  $N$ independent sites has a probability $P_{0}(l)$ of being broken and
$1-P_{0}(l)$ of being intact. The consequences of this distribution function 
 for the mechanical behavior and  correlation
length at the transition point of macroscopic rupture will be developed
in section \ref{sub:Global-load-model}.

\subsection{Local load sharing \label{sub:Local-load-sharing}}

\subsubsection{Damage-state probability distribution  
\label{sub:obtention-of-probability}}

We now consider the case where each broken cell increases the local
deformation by an amount $\alpha\varepsilon l$ on each of its nearest
neighbors;  i.e.,  $J_{xy}=\alpha$ for each nearest neighbor pair, or $J_{xy}=0$
for  more distant cells.

Note that for a given cell $x$, the potential energy release defined
in Eq.\ (\ref{eq:expr,delta,E,p}), $\Delta E_{p}[\varphi,x,l]$,
is a growing function of both the deformation $l$ and the subset
of cracked cells $\varphi$ considered -- in the sense that if we
consider two subsets $B\subset D$ and $x\in D^{C}$, 
then $\Delta E_{p}[\varphi^{D},x,l]\geq\Delta E_{p}[\varphi^{B},x,l]$.
Physically, this inequality means that the local load over each cell increases
with the external load imposed, and that any cell breaking anywhere
else induces  an additional increase in local load. 
This inequality will play a key role in obtaining 
 the damage-state  probability distribution. 

To aid the pedagogic development, we first derive the probability of occurence
of a configuration $\varphi^{B}$ with $n$ isolated broken cells forming
a subset $B$, which do not share any common nearest neighbors. There
are  $zn$ nearest neighbors to these $n$ cells (in a subset
$F$ corresponding to the boundary of broken cells)  where $z$ is the coordination
number of the lattice considered. 
There are then  $N-(z+1)n$ cells completely isolated from the broken
ones  in the subset $(B\cup F)^{C}$. For a cell $x$ having broken
from an intermediate stage $\varphi^{A}$ with $A\subset B$, at an
intermediate external load $h\in[0,l]$, the change in the stored energy satisfies  
 $\Delta E_{p}[\varphi^{A},x,h]=c h^{2}\leq c l^{2}$.  
For every cell having survived, 
we have either $\Delta E_{p}[\varphi^{B},x,l]=(c+\alpha\varepsilon) l^{2}$
if $x\in F$ ($x$ is  on the boundary of the broken cells set),  or
$\Delta E_{p}[\varphi^{B},x,l]=c l^{2}$ if  
$x\in(B\cup F)^{C}$ ($x$ is completely disconnected from the broken cells set). 
Applying  Griffith's principle [Eq.\ 
(\ref{eq:cond,break})]  to every cell
and intermediate deformations $h \in [0,l]$, 
 and using the monotony of $\Delta E_{p}[\varphi,x,l]$ in
both $l$ and $\varphi$, we obtain that a necessary and sufficient
condition for all cells in $B$ to be broken is that their surface
energy thresholds were below $c l^{2}$, while those of
their neighbors in $F$ were above $(c+\alpha\varepsilon) l^{2}$,
and the remaining ones in $(B\cup F)^{C}$ above $c l^{2}$.
Defining 
\begin{equation}
P_{1}(l)=P(e<(c+\alpha\varepsilon) l^{2}) 
= \int_0^{(c + \alpha \varepsilon) l^2} p(z) \, dz,\label{eq:P,1,of,x}\end{equation}
this then implies that 
\begin{equation}
P[\varphi,x]=P_{0}^{n}(1-P_{1})^{zn}(1-P_{0})^{N-(z+1)n}.
\label{eq:P,config,isolated,cracks}\end{equation}
is the probability of occurence of such a configuration.  
\begin{figure}[htbp]
\includegraphics[%
  width=0.90\columnwidth,
  keepaspectratio]{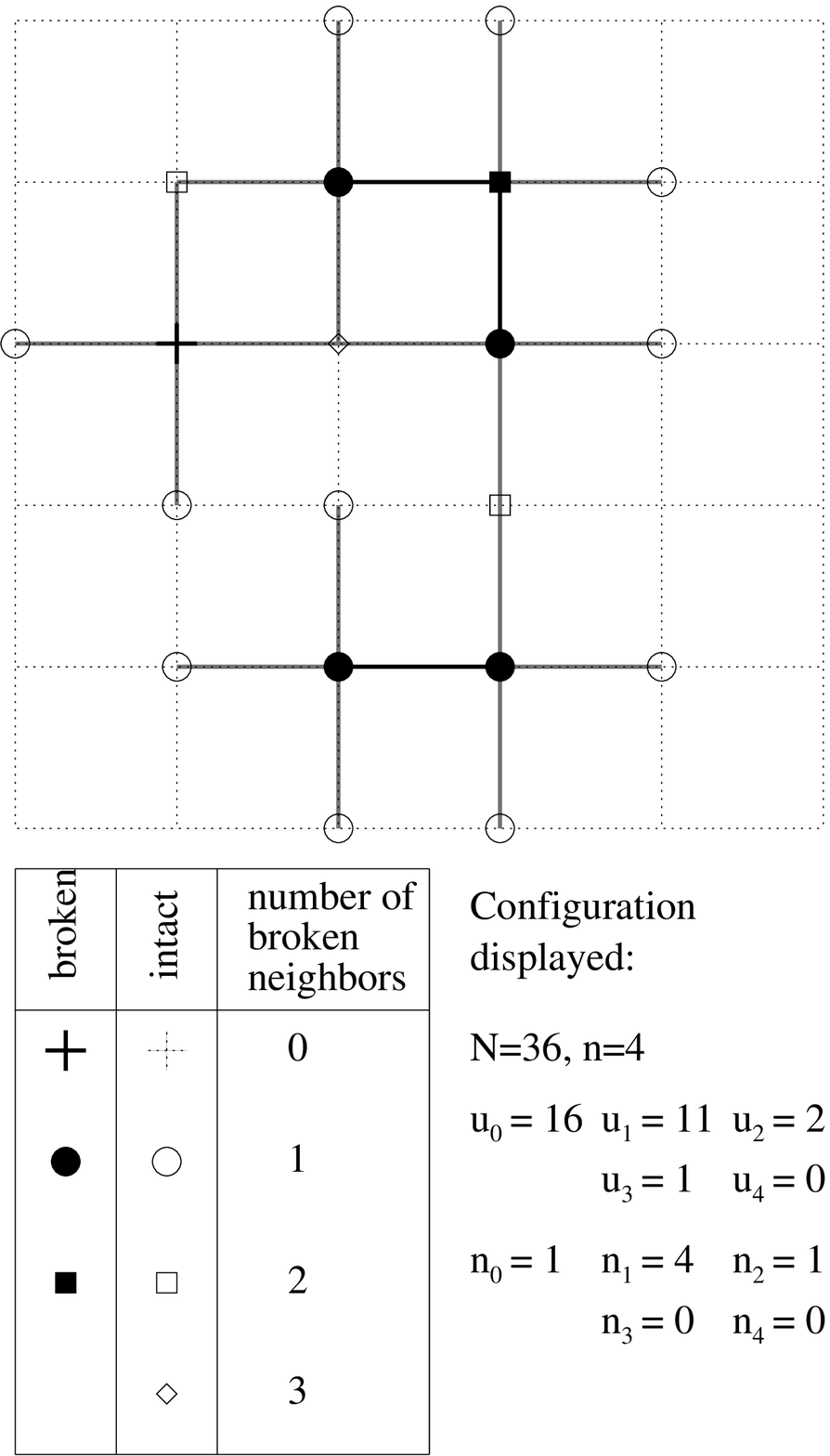}
\caption{Typical configuration and associated connectivity measures.\label{figure:machin}}
\end{figure}

We now pass to the  more general case.  In the argument, we obtain upper and  lower  bounds  
for the probability  
of some arbitrary damage state, and then demonstrate that in the limit 
 $\varepsilon/c \ll 1$, the 
two bounds  converge to the unique probability distribution of interest.

For any configuration $\varphi$,
 $u_{k}$ is defined as the number of intact cells with $k$ broken nearest neighbors and 
$n_{k}$ as the number of broken cells with $k$ broken neighbors.  If $n$ out 
of the $N$ cells are broken, we have  
 $\sum_{k}n_{k}=n$ and $\sum_{k}u_{k}=N-n$. 
The way  the above quantities are associated to any particular
configuration is illustrated in Fig.~\ref{figure:machin}.

 For a cell
$x$ which  broke from an intermediate state $\psi\subset\varphi$
with $k$ already broken neighbors in $\psi$, at an external load
$h$, we have 
\begin{equation}
\Delta E_{p}[\psi,x,h]=(c+k\alpha\varepsilon) h^{2}.  
\label{eq:Delta,E,p,with,k,broken,neighbs} 
\end{equation}
Using  again the monotony of $\Delta E_{p}$,  a necessary
and sufficient condition for any intact cell $y\in B^{C}$ to have
survived is that its threshold exceeded $\Delta E_{p}[\varphi,y,l]$.
The probability for each of these independent statistical events to
occur is expressed as $1-P_{k}(l)$, where 
\begin{equation}
P_{k}(l)=P(e<(c+k\alpha\varepsilon) l^{2}).\label{eq:P,k,of,x}\end{equation}
For any cell which broke $x\in B$, we note that they have broken
with certainty at the ultimate deformation $l$ if the external load was
sufficient to trigger their break without the help of 
overload due to  breaks of the other ones;  i.e., they have broken with 
certainty  if their energy
threshold was below $\Delta E_{p}(\varphi^{\emptyset},x,l)=c l^{2}$ where 
$\emptyset$ denotes the empty set (no broken cells).
Furthermore, if every threshold in $B$ was below $c l^{2}$,
except a particular one $x\in B$ which has $k_{x}$ broken neighbors
in the considered configuration $\varphi^{B}$ and has a breaking 
energy  between
$c l^{2}$ and $(c+k_{x}\alpha\varepsilon) l^{2}$,
the $k_{x}$ neighbors of this considered cell $x$ will have broken
with certainty at the ultimate load $l$, so that $x$ will also break
with certainty under the effect of the overload due to its broken
neighbors.  The probability that this individual threshold broke under
the sole effect of the overload due to its neighbors can be expressed
\begin{equation}
\Delta P_{k_{x}}  =  P_{k_{x}}(l)-P_{0}(l)\label{eq:def,delta,P,k}\end{equation}
Thus, a lower bound for the probability of occurence of the configuration
$\varphi^{B}$ can be expressed as 
\begin{eqnarray}
\lefteqn{P[\varphi^{B},l]>}\nonumber \\
 &  & \left[P_{0}^{n}+\left(\sum_{x\in B}P_{0}^{n-1}\Delta P_{k_{x}}\right)\right]
\prod_{k\in{N}}(1-P_{k})^{u_{k}}\label{eq:lower,bound,proba,config}\end{eqnarray}
where the index $k$ runs formally to ${N}$; however,   $u_{k}=0$
when $k>z$ (the coordination number of the lattice). 

For any cell $x\in B$ which broke, its  associated
energy threshold was necessarily lower than 
$\Delta E_{p}[\varphi^{B\smallsetminus\left\{ x\right\} },x,l]$ 
where $B\smallsetminus\left\{ x\right\}$ denotes the set $B$ with 
cell $\{x\}$ excluded from it.
Thus, an upper bound for the probability of occurence
of the configuration under study, is 
 \begin{eqnarray}
\lefteqn{P[\varphi^{B},l]<}\nonumber \\
 &  & \left[\prod_{x\in B}P_{k_{x}}\right]\prod_{k}(1-P_{k})^{u_{k}}=\nonumber \\
 &  & \left[\prod_{x\in B}(P_{0}+\Delta P_{k_{x}})\right]\prod_{k}(1-P_{k})^{u_{k}}.  
\label{eq:upper,bound,proba,config}\end{eqnarray}
For a continuous p.d.f.\ over thresholds (that guarantees   no jumps
in the cumulative distribution $P$ and is the only restriction placed on the p.d.f.),  
$\Delta P_{k}$ is a quantity of order $\varepsilon$, and $P_{0}$ is of order
$1$. In the limit of a Born model (weak stress perturbations for which 
 $\varepsilon/c\ll 1$), the upper and lower bounds for the probability
of occurence  are identical to order $\varepsilon$.
We have thus established in this framework that 
\begin{equation}
P[\varphi,l]=\prod_{m=0}^{\infty}P_{m}{}^{n_{m}}\prod_{k=0}^{\infty}(1-P_{k})^{u_{k}}.
\label{eq:expr,proba,config}\end{equation}

\subsubsection{Identification of a surface tension and cohesion energy 
\label{sub:Identification-of-surface}}

The above can be re-expressed for small interactions $\varepsilon\ll1$
by a Taylor expansion of the cumulative q.d.\ (quenched disorder) distribution
as 
\begin{equation}
\Delta P_{k}(l)  =  \gamma(l)k 
\mbox{\hskip3mm where \hskip3mm}  
\gamma  =   p(c l^{2}) \, \alpha \varepsilon l^{2}.  
\label{eq:def,gamma}
\end{equation}
Then, Eq.\ (\ref{eq:expr,proba,config}) 
becomes
\begin{eqnarray}
\lefteqn{\ln P[\varphi,l]=}\nonumber \\
 &  & \ln\left[\prod_{m=0}^{\infty}(P_{0}+\gamma m)^{n_{m}}
\prod_{k=0}^{\infty}(1-P_{0}-\gamma k)^{u_{k}}\right]\label{eq:expr,proba,config2}\\
 & \simeq & n\ln P_{0}+(N-n)\ln(1-P_{0})+\nonumber \\
 &  & \frac{\gamma}{P_{0}} \sum_{m=1}^{\infty}mn_{m}- 
\frac{\gamma}{1-P_{0}} \sum_{k=1}^{\infty}ku_{k}\nonumber \\
 & = & n\ln P_{0}+(N-n)\ln(1-P_{0})+\nonumber \\
 &  & 2\frac{\gamma}{P_{0}}n_{I}-\frac{\gamma}{1-P_{0}}n_{S}\label{eq:Taylor,exp,log,P,config}\end{eqnarray}
where in the state $\varphi$ considered, $n$ is the number of
broken sites, and $n_{I}$ and $n_{S}$ refer respectively to the
number of internal bonds between pairs of broken cells and the number
of boundary bonds between broken and intact cells.

These probabilities are thus of the form 
 \begin{equation}
P[\varphi,l]=P[\varphi^{\emptyset}]
e^{n\ln(P_{o}/(1-P_{0}))+2n_{I}{\gamma}/{P_{0}}-n_{S}
{\gamma}/{(1-P_{0})}}\label{eq:Boltzamnnian,local,int,model}
\end{equation}
with the probability of the intact state 
\begin{equation}
P[\varphi^{\emptyset}]=(1-P_{0})^{N}. \label{eq:P,ref,intact}
\end{equation}
The probabilities  are thus classic Boltzmann distributions 
$ P[\varphi^{\emptyset}] e^{-H}$ where 
 \begin{equation}
-{H[\varphi,l]}=\ln\left[\frac{P_{o}}{(1-P_{0})}\right]n+2\frac{\gamma}{P_{0}}n_{I}-
\frac{\gamma}{1-P_{0}}n_{S}.\label{eq:product,Hamiltonian,temp}\end{equation}
For no interactions, $\gamma=0$ and the above reduces to the global-load model  
of  Eq.\ (\ref{eq:P,phi,glob,noint}). Since $n$
and $n_{I}$ can be formally interpreted as volume integrals over
the interior of clusters of broken cells, and $n_{S}$ as a surface
integral along the boundary of these clusters, we can make the 
following analogies to the quantities of 
classical statistical physics:   
\begin{eqnarray}
\ln\left[P_{0}/(1-P_{0})\right] & \rightarrow & -\mu/T\,\mbox{\hskip2mm chemical potential}\label{eq:chem,pot}\\
2\frac{\gamma}{P_{0}} & \rightarrow & \frac{-e}{T}\,\mbox{\hskip 7mm 
 bulk energy}\label{eq:bulk,energ}\\
\frac{\gamma\ell}{1-P_{0}} & \rightarrow & \frac{\gamma_{S}}{T}\,\mbox{\hskip7mm 
surface tension}.\label{eq:surf,tens}\end{eqnarray}
The first term in Eq.\ (\ref{eq:product,Hamiltonian,temp}) 
accounts for the average energy required to break a cell,  
 the second term for an increase in the probability of finding
some clusters of connected cracks due to positive interactions between
them, and the third term  for a decrease in the probability of finding clusters
with a long interface between cracked and non-cracked region due
to the fact that intact cells along the boundary are more likely to
have broken from overloading from cracked neighbors, thus leading to even more
fractured states. 

In the beginning of the process, the two first terms dominate, and
to leading order \begin{equation}
-{H[\varphi,l]}=  \left(\ln P_{0}\right) n +2\frac{\gamma}{P_{0}} n_{I}.
\label{eq:domin,order,process,start,no,surf,tension}\end{equation}
 For example, defining $\delta=\lim_{y\sim0}[\ln P(y)/\ln y]$, so  
that $ P(y)\sim ay^{\delta}$, in the small deformation limit, one has: 
$P_{0}\sim a c^{\delta}l^{2\delta};\,\gamma\sim\alpha a\delta\varepsilon 
c^{\delta-1}l^{2\delta};\,\gamma/P_{0}\sim\alpha\delta\varepsilon c^{-1}$; 
and, in decreasing order of magnitude, $-H\sim\ln(a c^{\delta}l^{2\delta})n +
a c^{\delta}l^{2\delta}n +2\alpha\delta\varepsilon c^{-1} n_{I} 
-a\alpha\delta\varepsilon c^{-1}l^{2\delta}n_{S}$.

\subsection{Arbitrary-range interactions\label{sub:arbitrary-range-interactions}}

As long as screening effects are absent or neglected (as they are in the 
present model of weak interactions), 
the above arguments based on Griffith's principle and
 the monotony of $\Delta E_{p}$ in $\varphi$ and $l$, 
 extend directly to the case of arbitrary ranges of interactions. 

At external deformation
$l$, a cell $x$ has: (1)  broken with certainty if the external load
alone could break it; i.e., if its associated surface energy is lower
than $c l^{2}$;  and (2) is intact with certainty if 
the external load plus the load perturbation due to the broken cells
in the final configuration could not break it at final deformation; i.e., 
if its surface energy is higher than $(c+\varepsilon\sum_{y}J_{xy}\varphi_{y}^{B})l^{2}$.
Denoting 
\begin{eqnarray}
P_{0} & = & P(e<c l^{2})\label{eq:def,P0}\\
\Delta P_{x} & = & P(e<(c+\varepsilon\sum_{y}J_{xy}\varphi_{y}^{B}) l^{2})-P(e<c l^{2})\nonumber \\
 & = & p(cl^{2})\varepsilon l^{2}\sum_{y}J_{xy}\varphi_{y}^{B}\,,\label{eq:def,delta,P,x}\end{eqnarray}
 a necessary condition to end up at a certain configuration
$\varphi^{B}$ at deformation $l$ is that all surviving cells $x\in B^{C}$
in that configuration have their threshold above $(c+\varepsilon\sum_{y}J_{xy}\varphi_{y}^{B}) l^{2}$,
and all cells which broke $x\in B$ have their threshold below $(c+\varepsilon\sum_{y}J_{xy}\varphi_{y}^{B}) l^{2}$.
Thus, \begin{equation}
P[\varphi^{B},l]<\prod_{x\in B}(P_{0}+\Delta P_{x})
\prod_{z\in B^{C}}(1-P_{0}-\Delta P_{z})\label{eq:upper,bound}\end{equation}
provides an upperbound for the probabilities in the case of arbitrary ranges of interaction.

Conversely, a sufficient condition to end up in this configuration
is that all surviving cells have their threshold above 
$(c+\varepsilon\sum_{y}J_{xy}\varphi_{y}^{B}) l^{2}$,
and that the broken cells either have: (1) 
 all  their thresholds  below $c l^{2}$; or (2) 
all but one located at $x$ have such thresholds, the last
one breaking only due to the overload from the others; i.e., the 
last one  has its
threshold bounded by $c l^{2}<e<(c+\varepsilon\sum_{y}J_{xy}\varphi_{y}^{B}) l^{2}$.
This gives  a lower bound for the probability of the configuration
$\varphi^{B}$,
\begin{equation}
P[\varphi^{B},l]>\sum_{x\in B}(P_{0}+\Delta P_{x})\prod_{y\in B\smallsetminus
\left\{ x\right\} }P_{0}\prod_{z\in B^{C}}(1-P_{0}-\Delta P_{z})\,.\label{eq:lower,bound}\end{equation}
As earlier, both lower and upper bound coincide to order $\varepsilon$, so
that upon  Taylor expanding $\ln P[\varphi,l]$ to this order, we
again obtain the Boltzmannian $P[\varphi,l]  =  P[\varphi^{\emptyset}]e^{-H[\varphi,l]}$ with 
an intact probability given again by $P[\varphi^{\emptyset}]  =  (1-P_{0})^{N}$ and  
\begin{eqnarray}
-{H[\varphi,l]} & = & \ln[P_{0}/(1-P_{0})] n\nonumber \\
 &  & +\frac{p(c l^{2})\varepsilon l^{2}}{P_{0}}\sum_{xy}J_{xy}\varphi_{x}\varphi_{y}\nonumber \\
 &  & -\frac{p(c l^{2})\varepsilon l^{2}}{1-P_{0}}\sum J_{xy}\varphi_{x}(1-\varphi_{y}).  
\label{eq:Hamilt,general}
\end{eqnarray}
In the beginning of the process, the two first terms once again dominate
and \begin{eqnarray}
- {H[\varphi,l]} & = & \ln[P_{0}] n\nonumber \\
 &  & +\frac{p(c l^{2})\varepsilon l^{2}\sum_{xy}J_{xy}\varphi_{x}\varphi_{y}}{P_{0}}
\label{eq:Hamilt,general,asympt}\end{eqnarray}
This expresses the equivalence between this most general weakly interacting
damage model and an Ising model with generalized interaction rules.

\section{Equivalence  with a maximum entropy postulate}

\label{sec:Equivalence-of-the}


We now obtain these same probability distributions using the standard entropy maximization 
argument.  

It is convenient to introduce the index $j$ to denote each possible damage 
configuration $\varphi$.  We postulate  that the probability distribution function over
configurations $j$ 
 maximizes Shannon's
disorder  \cite{Shannon48}
\begin{equation}
S=-\sum_{j}p_{j}\ln p_{j}\label{eq:shannon}\end{equation}
subject to the constraints 
\begin{equation}
\sum_{j}p_{j}E_{j}=U;\,\sum p_{j}n_{j}=n;\,\sum_{j}p_{j}=1;\,\forall j\, l_{j}=l
\label{eq:constraints}\end{equation}
where $U$ is the total average energy that has been put into the system 
and $n$ is again the average number of broken cells.  
The validity of this maximization postulate will be directly demonstrated in 
what follows. 
 However,  independent of the formal demonstration,  
 one can anticipate that 
Shannon entropy should  be maximized since the initial quenched disorder 
in the breaking energies allows each possible damage configuration to be accessible. 
The constraints allow for the content of the Griffith principle 
and are what make certain    
emergent damage states more  probable than others.  

Throughout the
remainder of the paper,  $p_{j}$ denotes the probability of finding
a configuration $j$ over all possible realizations of the q.d., when
the system has been brought to average deformation $l$ starting from
an initially intact state.  The term $n_{j}$ now refers to the total number
of cracks in the configuration $j$, while $n$ is the statistical average
of $n_{j}$. A few classical results can directly be derived from these assumptions:
\begin{eqnarray}
p_{j} & = & \frac{e^{-\beta(E_{j}-\mu n_{j})}}{Z}\label{eq:p,j}\\
Z & \equiv & \sum_{j}e^{-\beta(E_{j}-\mu n_{j})}\nonumber 
\end{eqnarray}
where $\beta = \left.\partial S/\partial U\right|_{l,n}$ and  
$\beta \mu = -\left. \partial S/\partial n \right|_{l, U}$. 
From these one further has 
\begin{eqnarray} 
G & \equiv & -\ln(Z)/\beta\nonumber \\
G & = & U-S/\beta-\mu n\nonumber \\
dG & = & \tau dl-Sd(1/\beta)-nd\mu \nonumber \label{eq:dG}\\
dU & = & \tau dl+dS/\beta+\mu dn\label{eq:dU,stat}\end{eqnarray}
where \begin{equation}
\tau  =  \sum_{j}p_{j}\tau_{j}\label{eq:average,stress} 
\mbox{\hskip3mm and \hskip3mm} 
\tau_{j}  \equiv  \frac{dE_{j}}{dl}.  \end{equation}
The  thermodynamic parameters $\beta$ and $\mu$ are  obtained  here 
by comparing Eq.\ (\ref{eq:p,j}) to the earlier probabilities 
obtained by direct integration  
over the microstate space.

The p.d.f.\  over configurations is  a maximum of Shannon's
entropy under the above constraints, if and only if   
there are
two constants $(\beta,\mu)$ such that  \begin{equation}
\forall j,\, H_{j}-\beta(E_{j}-\mu n_{j})=\text{{constant}}\label{eq:morphism}\end{equation}
with $H_j$  given by Eq.\ (\ref{eq:Hamilt,general}).
From Eqs.\ (\ref{eq:E,pot})--(\ref{eq:comp,2}) and Eq.\ (\ref{eq:Hamilt,general}),
we have
\begin{eqnarray}
-H_{j} & = & \ln(\frac{P_{0}}{1-P_{0}})n_{j}+\nonumber \\
 &  & \frac{p(c l^{2})\varepsilon l^{2}}{P_{0}}\sum_{xy}J_{xy}\varphi_{x}\varphi_{y}\nonumber \\
 &  & -\frac{p(c l^{2})\varepsilon l^{2}}{1-P_{0}}\sum_{xy}J_{xy}\varphi_{x}(1-\varphi_{y})\nonumber \\
E_{j} & = & (N-c n_{j}-\varepsilon \sum_{xy}J_{xy}\varphi_{x}\varphi_{y})l^{2}\nonumber \\
\sum_{xy}J_{xy}\varphi_{x}(1-\varphi_{y}) & = & -\sum_{xy}J_{xy}\varphi_{x}\varphi_{y}+n_{j}\sum_{r}J_{r}\label{eq:decomp,surf,term}\end{eqnarray}
where  translational invariance of $J_{xy}=J_{\mathbf{r}=\mathbf{y}-\mathbf{x}}$ has been assumed. 
Equation (\ref{eq:morphism}) then requires 
\begin{eqnarray}
\beta & = & \frac{p(c l^{2})}{P_{0}(1-P_{0})}\label{eq:beta}\\
\mu & = & \frac{P_{0}(1-P_{0})}{p(c l^{2})}\ln\left[\frac{P_{0}}{(1-P_{0})}\right] - l^{2}( 
c +\varepsilon P_{0}\sum_{r}J_{r}).  \label{eq:mu}\end{eqnarray}
Thus, the  p.d.f.\ over
configurations is indeed maximizes  entropy [Eq.\ (\ref{eq:shannon})] 
under the constraints of Eq.\ (\ref{eq:constraints}), with no unknowns.  
The inverse temperature and chemical potential depend on the deformation
level  through Eqs.\ (\ref{eq:beta})--(\ref{eq:mu}).  They   are well defined
analytical functions of $l$ and the model parameters considered,
the q.d.\ distribution via $P_{0}$,  and the interaction coupling $J_r$. So the usual
machinery of equilibrium statistical mechanics [Eqs. (\ref{eq:p,j})--(\ref{eq:dG})],
is valid and can be used for any of our damage models. 

The autocorrelation
function $\langle\varphi_{x}\varphi_{y}\rangle$ can therefore be obtained in the
standard way by: (1)  defining a new Hamiltonian $E_{j}'=E_{j}+\sum_{x}Q_{x}\varphi_{x}$ 
that incorporates a 
 coupling with a formal external field $Q_{x}$; (2) evaluating
the associated generalized partition function $Z'$;  and (3) performing
second-order derivatives with respect to the external field, 
\begin{equation}
\left\langle \varphi_{x}\varphi_{y}\right\rangle =\frac{1}{\beta^{2}Z'}
\left.\frac{\partial^{2}Z'}{\partial Q_{x}\partial Q_{y}}\right|_{Q\equiv0}.  
\label{eq:principle,autocorrel,funct}
\end{equation}
 Second order derivatives of the free energy $G$ with respect to
$\beta$ and $\mu$ can also be directly  related to variances of the
number of broken cracks $n_{j}$ and the potential elastic energy $E_{j}$,
but these standard derivations will be left to the attention of the
reader.

Note that the formal temperature $1/\beta$ (the energy yardstick used to distinguish   
 probable from improbable states) behaves regularly throughout
the damage process (for continuous q.d.\ distributions).   In the case of a
uniform q.d.\ on {[}0,1{]}, it will take the particularly 
simple form $1/\beta=P_{0}(1-P_{0})$,
starting from zero and going back to it, with a maximum when $P_{0}=1/2$. 
The chemical potential
$\mu$ behaves regularly as well.

\section{Applications}
\label{sec:Applications}
\subsection{Global load model\label{sub:Global-load-model}}

A consistency check of the results in 
 Section
\ref{sec:Equivalence-of-the} is now performed for the case of the simple 
global-load sharing model.  
From Eqs.\ (\ref{eq:beta})--(\ref{eq:mu})
with $J_{r}=0$, we have
\begin{eqnarray}
E_{j}&=&(N-c n_{j})l^{2}\label{eq:E,noint}\\
\beta(E_{j}-\mu n_{j}) &=&  -  \ln\left[\frac{P_{0}}{1-P_{0}}\right]n_{j}\nonumber  \\
 &  & +\frac{p(c l^{2})}{P_{0}(1-P_{0})}[(N-c n_{j})l^{2}+c n_{j}l^{2}]\nonumber \\
 & = & \ln\left[\frac{P_{0}}{1-P_{0}}\right]n_{j} +f(l).   \label{ident,pdfs} 
\end{eqnarray}
Independently, we also have the direct result 
 from Eq.\ (\ref{eq:P,phi,glob,noint}) 
\begin{equation}
p_{j}=P_{0}^{n_{j}}(1-P_{0})^{N-n_{j}}
=P_{0} e^{-n_{j} \ln[P_0/(1-P_0)]}.
\label{eq:p,j,noint}
\end{equation}
Thus, the Boltzmann distribution Eq.\ (\ref{eq:p,j}) with temperature
and chemical potential given by Eqs. (\ref{eq:beta}) and (\ref{eq:mu})
is indeed identical to the known solution Eq.\ (\ref{eq:p,j,noint}),
which confirms the validity of the expressions for $\beta$ and $\mu$.

It is also instructive to look at all terms in the first law Eq.\ (\ref{eq:dU,stat})
to see exactly what they  represent.  
The values of the average quantities
in this simplest model can be obtained using the lemma 
$$\sum_{j}P_{0}^{n_{j}}(1-P_{0})^{N-n_{j}}n_{j} \label{lemmaB} 
=N P_{0}$$
which is demonstrated by applying  the operator $x\partial/\partial x$  to
the binomial theorem
$$(x+y)^{N}=\sum_{n=0}^{N}\frac{N!}{(N-n)!n!}x^{n}y^{N-n}=\sum_{j}n_{j}x^{n_{j}}y^{N-n_{j}},$$
and then taking $x=P_{0}$ and $y=1-P_{0}$.  
Using Eq.\ (\ref{lemmaB}), one obtains  
\begin{eqnarray}
n & = & P_{0}N \nonumber \\
U & = & \sum_{j}P_{0}^{n_{j}}(1-P_{0})^{N-n_{j}}(N-c n_{j})l^{2}\nonumber \\
 & = & N(1-c P_{0})l^{2}\nonumber \\
S & = & -N[P_{0}\ln P_{0}+(1-P_{0})\ln(1-P_{0})]\nonumber \\
\tau_{j} & = & \frac{dE_{j}}{dl}=(N-c n_{j})2l\nonumber \\
\tau & = & N(1-c P_{0})2l. \nonumber \end{eqnarray}
Taking the derivatives of these quantities yields   
\begin{eqnarray}
dn & = & N dP_{0}\label{eq:dmean1}\\
dS & = & \ln\left[\frac{P_{0}}{1-P_{0}}\right]dP_{0}\nonumber \\
\mu dn & = & \left(\frac{P_{0}(1-P_{0})}{p(c l^{2})}\ln\left[\frac{P_{0}}{1-P_{0}}\right]
-c l^{2}\right)N dP_{0}\nonumber \\
\frac{dS}{\beta} & = & -\frac{P_{0}(1-P_{0})}{p(c l^{2})}\ln\left[\frac{P_{0}}{1-P_{0}} 
\right] N 
dP_{0}\nonumber \\
\mu dn+\frac{dS}{\beta} & = & -c l^{2}N dP_{0} \nonumber \\
\tau dl & = & N (1-c P_{0})2ldl\nonumber \\
\mu dn+\frac{dS}{\beta}+\tau dl & = & N (1-c P_{0})2ldl-c l^{2}N 
dP_{0}\nonumber \\
 & = & d[N (1-c P_{0})l^{2}]=dU\nonumber \end{eqnarray}
which is a consistency check for the validity of the first law [Eq.\
(\ref{eq:dU,stat})].
The average mechanical behavior of this model, as well as the evolution
of entropy and formal temperature are illustrated in Fig.\ \ref{fig:averages,gls} 
for flat q.d.\ distributions between $0$ and $l_{\max}$ and for various 
values of $c$ (the parameter that controls the relative change in stiffness 
due to a cell breaking).  

\begin{figure}[h!]
\includegraphics[%
  width=0.85\columnwidth,
  keepaspectratio]{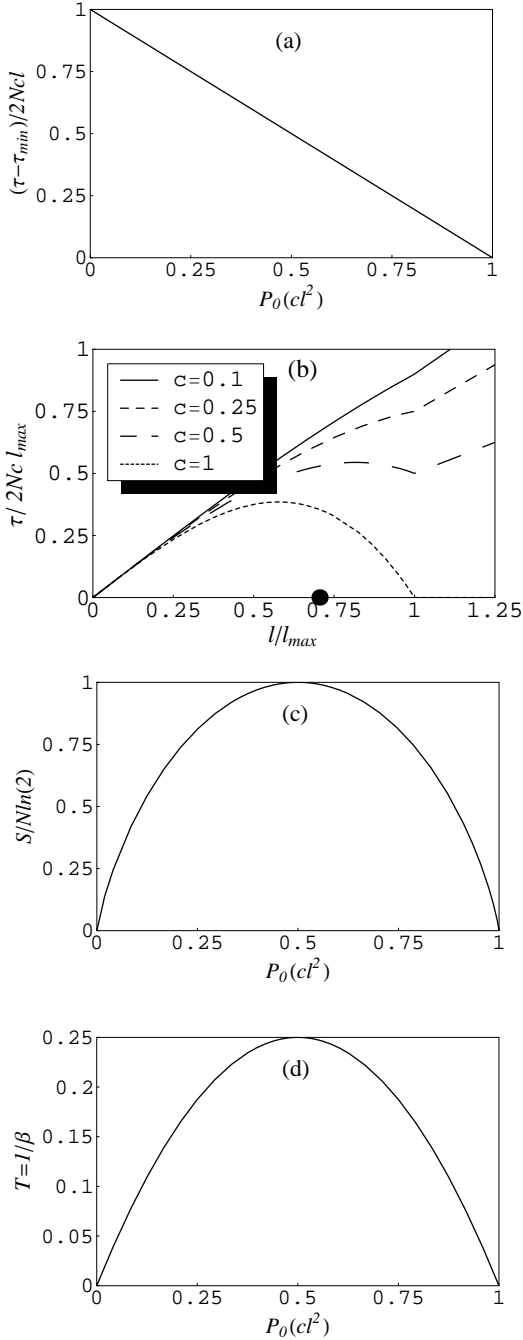}
\caption{Thermodynamic and mechanical response as a function of imposed 
deformation for the global-load sharing model:  (a) the difference between 
the average Young's modulus $\tau/l$  and the minimum Young's modulus 
$\tau_{min}/l$ (that holds in the entirely damaged configuration); 
(b) average stress for a 
few particular models (the black dot is the percolation critical-point transition); 
(c) Shannon entropy; and (d) temperature.
\label{fig:averages,gls}}
\end{figure}

Most importantly for our present purposes, 
since the probabity $P_0$ of having a site broken in this
model is  independent of the configuration and site location,
the global load-sharing damage  model is exactly equivalent 
to the percolation model with occupation
probability $P_{0}(l)$. There is a critical-point 
phase transition in this model,
when $P_{0}(l_{c})=1/2$, for which $S$ goes through a maximum $S=N\ln2$.
The correlation length diverges then as 
\begin{equation}
\xi\sim|P_{0}(l)-1/2|^{-\nu}\label{eq:xi,p,MF}\end{equation}
with $\nu=4/3$ in dimension $D=2$ \cite{StaufferA94}. We have in
general \begin{equation}
P_{0}(l)-P_{0}(l_{c})\sim p(c l_{c}^{2})(l-l_{c})\label{eq:P,l,MF}\end{equation}
and thus \begin{equation}
\xi\sim|l-l_{c}|^{-4/3}.\label{eq:xi,l,MF}\end{equation}
In pathological cases,  special q.d.\ distributions satisfy $p(c l_{c}^2)=0$
so that  $P_{0}(l)-P_{0}(l_{c})\sim(l-l_{c})^{\alpha}$ with $\alpha\neq1$.
This results in 
\begin{equation}
\xi\sim|l-l_{c}|^{-4\alpha/3}.\label{eq:xi,l,patho}\end{equation}
Note that in Ref.\cite{PrideT02}, we have treated this model with
$c=1$, which did not change anything to the nature of the
transition, but only changed the minimum stiffness associated with 
the most damaged configuration where 
$\tau_{min}=2N(1-c)l$,   and, consequently,  the position
or existence of a peak stress in the average mechanical response $\tau(l)$.
This is seen in  Fig.~\ref{fig:averages,gls}(b): the existence,
and the position of a possible peak-stress position relative to the
percolation transition (black dot $l/l_{max}=1/\sqrt{2}$),
depends on the particular model considered. However, the divergence
of the correlation length and the associated critical-point nature of the percolation
transition  are insensitive to  $c$.  

The present approach  does not allow a direct exploration of  the avalanche
distributions as the critical point is approached.  
This is because 
 the probability distribution over configurations was obtained
at each elongation $l$  by averaging  over all realizations
of the quenched disorder. To obtain directly a result on avalanches,
on the contrary, correlations between successive elongations should be
considered for each realization of the q.d., and the average over q.d.\ 
should only be considered afterwards. Our damage  model  nonetheless 
 behaves as a fiber bundle  with GLS for which results about 
the avalanche distribution have been determined \cite{Kloster97,HansenHemmer94,HemmerHansen92}.
If $c$ is sufficiently large  that the average strain/stress
curve has a peak at $\tau_{c}$, there is a burst in  the number
of elements that must break  before the system is able to sustain the same
global force again.   The distribution of the burst size $\Delta$ should
thus scale as \[
n(\Delta)\sim\Delta^{-5/2}e^{-\Delta/\Delta_{0}},\]
with $\Delta_{0}\sim|\tau-\tau_{c}|^{-1/2}$. It might be tempting
to relate this power-law behavior of the burst distribution to the
behavior of cluster numbers in mean-field percolation (on a Bethe lattice
or on hypercubes in dimensionality $D\geq6$). The average number
of clusters of size $s$ per lattice site  also scales as 
$n_{p}(\Delta)\sim\Delta^{-5/2}e^{-\Delta/\Delta_{0}}$ in such models 
where $\Delta_{0}\sim|p-p_{c}|^{-2}$ \cite{StaufferA94}. But the
fact that both the 
cluster number distribution at critical percolation threshold and the  burst
size distribution at peak stress have the same power law exponent of -5/2 
might  simply be coincidental.   These
two points (critical percolation threshold and possible peak stress)
in general differ, and the exponents characterizing the evolution
of the upper cutoff $\Delta_{0}$ of these distributions close to
percolation threshold or peak stress seem also unrelated. 

\subsection{Local load model\label{sub:Local-load-model}}

In this case, $J_{xy}=\alpha$ when $x,y$ are nearest neighbors,
and zero otherwise. Defining the order parameter $\sigma\in\left\{ -1,1\right\} $
as $\sigma=2\varphi-1$, we note that\begin{equation}
\sum_{xy}J_{xy}\varphi_{x}\varphi_{y}=\alpha N 
+2\alpha\sum_{x}\sigma_{x}+\frac{\alpha}{4}\sum_{<xy>}\sigma_{x}\sigma_{y}\label{lemma}\end{equation}
where $<xy>$ denotes a sum on nearest neighbors only. Consequently,
the probability over configurations can be cast under the form
\begin{eqnarray}
P[\sigma,l] & = & e^{\beta(E\sum_{x}\sigma_{x}+J\sum_{<xy>}\sigma_{x}\sigma_{y})}/Z,
\label{eq:proba.LLS}\end{eqnarray}
which is exactly a classical Ising model with coupling constant and
external field given by
 \begin{eqnarray}
\beta J & = & \frac{\alpha p(c l^{2})\varepsilon l^{2}}{4P_{0}(1-P_{0})},\label{value,betaJ,LLS}\\
\beta E & = & \frac{1}{2}\ln\left[\frac{P_{0}}{1-P_{0}}\right]+
\frac{2\alpha p(c l^{2})\varepsilon l^{2}}{P_{0}}.  \label{value,betaE1}\end{eqnarray}
The critical point of this model is at \cite{Huang87}
 $$(\beta J_{c},\beta E)=(A_{c},0)$$
with $A_{c}=\ln(1+\sqrt{2})/2$. The external field
$\beta E$ starts at infinitely negative values, and ends up at infinitely
positive  ones. It evolves continuously and thus necessarily crosses
$E=0$ at the $l_{c}$ satisfying , from Eq.~(\ref{value,betaE1}),
\begin{equation}
P_{0}(l_{c})\left\{ \ln[1-P_{0}(l_{c})]-\ln[P_{0}(l_{c})]\right\} 
=4\alpha p(c l_{c}^{2})\varepsilon l_{c}^{2}. 
\label{eq:cond,for,l,c}\end{equation}
The mean-field percolation result of  $P_{0}(l_{c})=1/2$
is thus recovered when 
the coupling vanishes ($J=0$), which is a consistency check. 

More generally, for non-zero nearest coupling constants $\alpha$,
the system will undergo a first-order transition if at $l_{c}$
satisfying Eq.~(\ref{eq:cond,for,l,c}), the formal inverse temperature
satisfies \begin{eqnarray}
\lefteqn{{\beta(l_{c})J=\frac{\alpha p(c l_{c}^{2})\varepsilon l_{c}^{2}}{4P_{0}(l_{c})(1-P_{0}(l_{c}))}=}}\nonumber \\
 &  & \frac{\ln[1-P_{0}(l_{c})]-\ln[P_{0}(l_{c})]}{16(1-P_{0}(l_{c}))}>A_{c}.
\label{eq:cond,1st,order,transition}\end{eqnarray}
Depending on the value of $\beta(l_{c})J$, the system can display
 four types of behavior, that are schematically depicted  in Fig. \ref{cap:Paths-associated-to}:

\begin{figure}[htbp]
\includegraphics[%
  width=0.90\columnwidth,
  keepaspectratio]{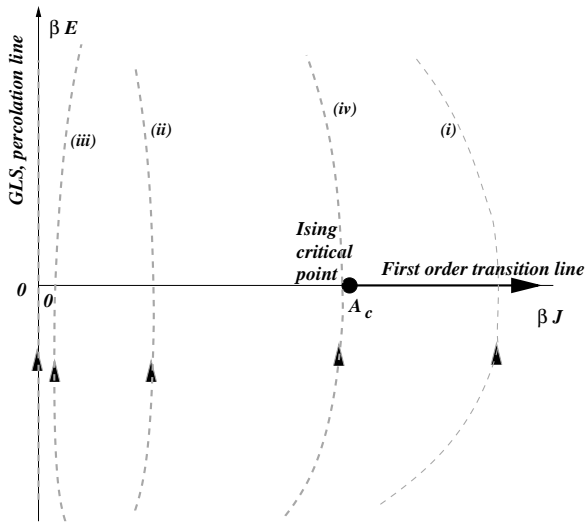}

\caption{Possible Paths in the space of coupling constants of Ising models,
under increasing imposed elongation associated to different Local
Load Sharing damage models.\label{cap:Paths-associated-to}}
\end{figure}

(i) For many q.d.\ distributions, the first value of $l_{c}$ satisfying
Eq.\ (\ref{eq:cond,for,l,c}) occurs for  very small values of
$P_{0}(l_{c})$, that correspond to small values of $2\alpha p(c l_{c}^{2})\varepsilon l_{c}^{2}$, 
 since $x\ln[(1-x)/x]\rightarrow0^{+}$ when $x\rightarrow0$. In
this case, $\ln[(1-P_{0}(l_{c}))/P_{0}(l_{c})]/(1-P_{0}(l_{c}))\simeq-\ln[P_{0}(l_{c})]\geq1$,
and the condition of Eq.~(\ref{eq:cond,1st,order,transition}) is
fulfilled.  The system thus goes through a first-order phase transition
at this $l_{c}$ and there is  a discontinuous jump in the average number
of broken cells and the average stress (which are related to
first derivatives of the free energy with respect to $l$ or $\mu$,
and are similar to the average number of spins up in an Ising model) \cite{Huang87}.
The correlation length increases up to the transition but remains
finite. All of this  behavior for  such local
load models has been documented in the literature  \cite{Kloster97,HansenHemmer94,Hidalgoetal02}. 

However,  three other behaviors are also possible if 
$\beta(l_{c})J<A_{c}$ at the  point $l_c$ at which 
the external field vanishes 
[Eq.\ (\ref{eq:cond,for,l,c})].  Whether this occurs is controlled by 
the type of q.d.\ distribution and the value of the coupling constant
$\alpha$.  When  $\beta(l_{c})J<A_{c}$, 
 no first order transition is encountered and  one can further 
classify the transition  into three subcategories:

(ii) If $\beta(l_{c})$ has a finite value of order unity, significantly
below $A_{c}$, the system simply goes continuously through $l_{c}$,
without discontinuity in sustained load or average number of cracks.
The correlation length remains finite.

(iii) If $\beta(l_{c})J\ll 1$, (which should happen for vanishing
$\alpha$), the distribution over configurations is dominated by the
external field, and the system essentially behaves as a percolation
model going through the percolation transition. Although there should
be corrections due to the nonzero character of $\beta(l_{c})J$, these
might be smaller than the finite-size corrections in numerical realizations,
and the correlation length would then be found to diverge up to the
system size as $\xi\sim|l-l_{c}|^{-4/3}$. 

(iv) Finally, if $\beta(l_{c})J\sim A_{c}^{-}$, the system comes
close to the critical point of the Ising model at $l_{c}$, and the
correlation length should diverge corresponding as in an Ising system
whose temperature comes close to $1/A_{c}=T_{c}$ from above, while
the external field reverts sign. The slope of the average mechanical
curve $\tau(l)$ should also locally diverge around $l_{c}$. The
exponents associated with the divergence of $\xi$ as function of
$(l-l_{c})$ depend on the way the critical point is approached as a 
function of $l$. For values of $l$ such that $\beta E(l)\ll 1$, we
write $1/\beta(l)J\sim T_{c}+f(l)$, and $\xi\sim|T(l)-T_{c}|^{-\nu}=|f(l)|^{-\nu}$
where $\nu=1$ for the 2D Ising model. The correlation 
length   therefore diverges  as  
 $|l-l_{c}|^{-\nu}$, unless the temperature has a quadratic minimum 
in $l$ close to $l_c$ in which case it diverges as  $|l-l_{c}|^{-2\nu}$.  

\subsection{Power-law decay\label{sub:Power-law-decay}}

We last consider the general case of stress perturbations decaying
as power-laws of the distance to broken cells, $J(x)\sim\alpha(x/d)^{-\gamma}/\Phi(\gamma)$,
where $\alpha$ is of order unity, $d$ is the lattice constant, and
$\Phi(\gamma)=\sum_{y\neq x}(|y-x|/d)^{-\gamma}$ is a normalizing
factor (which depends on the lattice size if $\gamma>D$, where $D$
is the system´s dimension). To
be consistent, the models considered here are biperiodic of linear
size $L$, and the interactions are put to $0$ for distances above
$L$. This type of model, considered for example by Hidalgo et al
\cite{Hidalgoetal02,Yewandeetal02}, allows us to span ranges between
purely global sharing (when $\gamma=0$), to the local sharing limit
($\gamma\rightarrow\infty$). Equations (\ref{eq:decomp,surf,term}-\ref{eq:mu})
show that this model leads to probability distributions over configurations
of the form
\begin{eqnarray}
P[\sigma,l] & = & e^{\beta(E\sum_{x}\sigma_{x}+\sum_{x\neq y}J(|x-y|)\sigma_{x}\sigma_{y})}/Z,\label{eq:proba.PLLS}\end{eqnarray}
which is a generalized long-range Ising model with coupling constants
and external field given by: \begin{eqnarray}
\beta J(r) & = & \frac{\alpha p(c l^{2})\varepsilon l^{2}}{4P_{0}(1-P_{0})}\frac{(r/d)^{-\gamma}}{\Phi(\gamma)},\label{value,betaJ,PLLS}\\
\beta E & = & \frac{1}{2}\ln(\frac{P_{0}}{1-P_{0}})+
\frac{\alpha p(c l^{2})\varepsilon l^{2}}{2P_{0}}.  \label{value,betaE}\end{eqnarray}
Note that although the case $\gamma\rightarrow\infty$ is isomorphic
to the Local Load Sharing Model introduced in the previous section,
the Global Load Sharing Model of Section \ref{sub:Global-load-model}
corresponds to $\alpha=0$, but not to the exponent $\gamma=0$. The presence
of the quadratic coupling makes this later case isomorphic to a Curie-Weiss
model, which is the Mean-Field theory of Ising models.  

Although such
long-range Ising models are still an open area of research, it has
been proposed \cite{CannasGT03,CannasMT00,CannasT96} that their behavior
can be classified into two categories depending on whether $\gamma>D$
or $\gamma<D$.   In the short-range case, $\gamma>D$, this model admits
a traditional thermodynamic limit when $L\rightarrow\infty$.
 Although $\Phi(\gamma)$ diverges
when $L\rightarrow\infty$, the thermodynamic limit is well defined
once $\Phi(\gamma)$ is introduced  into  the coupling constants 
$J(r)={(r/d)^{-\gamma}}/{\Phi(\gamma)}$.  Thus, the free or internal
energy per lattice site, entropy and  magnetization all 
admit a finite limit
when $L\rightarrow\infty$, and are functions of $\beta$ and the external
field $E$. 

In our case, we have 
\begin{eqnarray*}
\beta&=&\frac{\alpha p(c l^{2})
\varepsilon l^{2}}{4P_{0}(1-P_{0})} \\
\beta  E &=&\frac{1}{2}
\ln(\frac{P_{0}}{1-P_{0}})+\frac{\alpha p(c l^{2})\varepsilon l^{2}}{2P_{0}}.
\end{eqnarray*}
With $J$ defined as above, the Ising model shows that in the limit
$\gamma\rightarrow\infty$, the critical point is at \[
\lim_{\gamma \rightarrow \infty} \beta_{c}=\ln(1+\sqrt{2})/2.\]
It has also been shown \cite{CannasT96} that as $\gamma\rightarrow D^+$, 
$ \beta_{c}\sim 1$.
All short-range models $\gamma>D$ display critical points at finite
temperature, and since all of them have a finite range of interaction,
one can conjecture that all of them belong to the same universality
class as the Ising model, with a critical point such that  $\beta_{c}$
is of order unity \footnote{More complicated scenarii are also plausible, as illustrated by the behavior of spherical models depending on D and $\gamma>D$, see Domb, {\em The critical point}, (Taylor and Francis, London), p.189 for a review.}.
It has also been conjectured that all nonlocal models $\gamma<D$
belong to the Curie-Weiss  universality class of the Mean Field Ising model.
They also present critical temperatures such that  $\beta_{c}$
is of order one \cite{CannasMT00}.
The Curie-Weiss model corresponds to the mean-field coupling $J_{xy}=1/N$, 
independently of $x$ and $y$, where $N$ is total number of cells in the system.
For the Curie-Weiss model, the p.d.f.\ over state configurations is 
$P[\sigma] \propto \exp[\beta(E\sum_x \sigma_x + \sum_{xy} \sigma_x \sigma_y /N)] 
\propto \exp[\beta(E \sum_x \sigma_x + (\sum_x \sigma_x)^2 /2N)]$. 

All this suggests the following  classifications of  our  damage models:

(1)
If the coupling constant is
small ($\alpha\ll 1$),  the damage model is close to the percolation
model so that in the approach to the
transition at elongation $l_{c}$, the correlation
length behaves as $\xi\sim|l-l_{c}|^{-\nu}$, with $\nu=4/3$;

(2) For non-negligeable coupling constants $\alpha$, in the short-range case $\gamma>D$, we
recover the same three possible behaviors  as described above for
the local load sharing
rule;

(3)  In the long-range case where $\gamma<D$, 
we again recover the same type of four scenarios depending on the ratio 
$\beta(l_c)/A_{c}$ (as discussed in the local-load sharing case).
 The model behaves as percolation when  
$\alpha \ll 1$.  Otherwise, the  behavior  
is determined by the ratio $\beta(l_c)/A_{c}$. Note that $A_{c}$ is of order unity,
but depends on the particular exponent $\gamma$ and on the system
size. If $\beta(l_c)/A_{c} < 1$, the system behaves continuously and no transition is observed. 
If $\beta(l_c)/A_{c} > 1$, there is a discrete jump in both the average number of broken cells  
and the sustained load. The correlation length remains finite in both of these two cases. 
Only the limiting case of $\beta(l_c)/A_{c}\sim 1$  corresponds to a second-order phase transition.
In any of these cases, for large enough systems, the Curie-Weiss description holds according to 
Refs.\ \cite{CannasGT03,CannasMT00,CannasT96}.  
Accordingly, if there is any divergence of correlation length due to the system coming close to  
$\beta(l_c)/A_{c} \sim 1$ at $E=0$, the associated exponents should be those of the 
Curie-Weiss mean-field critical point ($\nu=1/2$), and not the Ising one. 

These results can be compared to numerical-simulation results in the literature for
related models. In fiber bundle models with power-law  interactions
\cite{Hidalgoetal02}, a transition has been found as a function of
the interaction exponent $\gamma$  that
is consistent with the above
analysis, predicting mean-field behavior for the long-range
case $\gamma<D$, and Ising-like behavior in the short-range case $\gamma>D$. Typical configurations prior
to breakdown for this type of system are displayed in Fig.\ 5 of Ref.\
\cite{Hidalgoetal02}, and look very similar to percolation configurations
close to the transition in the case $\gamma=0$,   displaying shorter and shorter  cluster sizes (characteristic  of the autocorrelation
length) and compact configurations as $\gamma$ increases above $D$. This is
coherent with a mean-field behavior close to percolation transition in the first case, as opposed
to a first order transition (analogous to Ising model crossing $E=0$
below $T_{c}$) when $\gamma<D$. This analogy is even more apparent
in Fig.\ 7, bottom, of Ref. \cite{Yewandeetal02}, where an extension
of this model was considered, with time-delayed fiber breaking process
in addition to this power-law decaying interactions \cite{Yewandeetal02}.

Burned fuse models, in which the interactions between burned
fuses decay as $1/r^{D}$, exhibit diverging autocorrelation lengths
at breakdown, with $\xi\sim|l-l_{c}|^{-\nu}$ where $l$ is the
imposed voltage, and $\nu$ is equal to the percolation exponent
\cite{StaufferA94}, $\nu=4/3$ in 2D  \cite{Hansen03}  or $0.88$ in 3D  
 \cite{Ramstad04}. The morphology of the connected ``fracture'' 
at breakdown is oriented, and different from percolating clusters
in the percolation model.  This seems to be  related to the anisotropic character
of  interactions in the  burned-fuse  model; i.e.,  the current
perturbation from a burned fuse goes as a dipolar field  decaying 
as $1/r^{D}$, but  also having an orientational aspect not included in
the models under study in this paper. Based on this, Hansen et al. \cite{Hansen03} 
relate the roughness of the spanning fracture to the autocorrelation
length exponent, based on arguments of percolation in gradient, and this  
 properly predicts the Hurst exponent of the final damage fronts, both in
2 and 3D. This anisotropic aspect is absent from the models discussed
here, but the fact that the autocorrelation length exponent is similar
to the ones of percolation, is coherent with the fact that long-range
systems are in the percolation universality class. Burned
fuse systems  are at the verge between short and long range interactions, in
the sense that they correspond to $\gamma=D$.

Last, fiber bundles connected to elastic plates, where $\gamma=1$
and $D=2$, have been numerically studied in Ref. \cite{Schmittbuhl03} and 
an autocorrelation length exponent of $\nu\simeq1.54$ numerically determined. 
The present theory
does not explain this autocorrelation exponent. The discrepancy between
this result and the percolation or Curie-Weiss critical point result, might presumably result
 from finite-size effects making this model ($\gamma=1$) still significantly different from the Curie Weiss one ($\gamma=0$), or from 
the fact that the stress perturbation was in this numerical model
too large for the small perturbation expansion performed here to apply.

\section{Conclusions}

\label{sec:Discussion-and-prospects}

We have treated a class of damage models having weak isotropic interactions 
between   cells that become damaged in the lattice.   A quenched
disorder is present in the rupture energies for each lattice
cell, and the  evolution of damage is ruled by the Griffith principle. 
Averaging over all possible realizations of the underlying
quenched disorder,  the probability distribution of each possible 
damage configuration was obtained as a 
 function of the deformation externally applied to the system. 

The exact calculation is analytically tractable in the case of a global
load sharing model, and it has been shown to be isomorphic to a percolation
model. This corresponds to the behavior of a system totally dominated
by the underlying disorder, where the next cell to break is always
the weakest one. Spatial interactions added to the system modify this
picture,  since the overload created by broken cells induce some degree of  
 spatial ordering that  competes  with the weakest-cell mechanism.
By limiting ourselves to small overloads compared to the average load of the system,
it was possible to obtain  the probability of damage
configurations  as  integrated over all realizations of the quenched disorder.

In this weak interaction limit, the resulting probability distributions were 
shown to be  
 Boltzmannians in the number of broken cells and in the stored
elastic energy. This type of distribution maximizes
Shannon´s entropy under constraints related to the energetic balance
of the fracture process, and we have demonstrated the formal relationship
between our quenched-disorder damage models and the 
 standard distributions arising in equilibrium
statistical mechanics.  This then allows the standard 
toolbox of statistical mechanics to be applied to our damage models.   

Our systems map onto three types of possible behaviors: 
(1)  percolation models in the case of interactions so weak,   
they may be neglected;  
(2)  Ising models for  non-negligeable short-range interactions; and 
(3) Curie-Weiss mean-field theory for non-negligeable 
long-range interactions.   The temperature
and external field in the partition function of our models 
are analytical functions that depend on the
 particular sharing rule,   on the type of quenched disorder considered,
and  on the average elongation (or deformation) externally loaded onto the system. 
The path followed in the Ising control parameter space when the load is increased from
$0$ depends on the q.d.\ distribution  and the load-sharing rule. When the formal external
field reverts sign, a phase transition is possible. This can correspond
to a first-order phase transition, a percolation-like transition,
or an  Ising critical-point transition, depending
on the value of the formal temperature during the transition.

The systems studied here are limited to  isotropic load perturbations.
We have earlier studied oriented crack models in \cite{ToussaintPr02a,ToussaintPr02b,ToussaintPr02c},
which corresponds to anisotropic load perturbations that  depend on
the orinetation of the crack opened in the lattice. Those earlier studies were 
 based on an entropy-maximum assumption. The hypotheses of the present work
extend directly to oriented systems, and so  the present paper  justifies
the entropy-maximum assumption postulated in our earlier work. The
precise value of the temperature, and the physical interpretation
of the functional forms given in \cite{ToussaintPr02a,ToussaintPr02b,ToussaintPr02c},
 should be modified
according to the results of this paper. Such modifications will, 
however,  result in identical functional forms relating the configuration
space and the p.d.f.\ over configurations, and thus the present work
confirms the existence of a phase transition in such an oriented crack
model, with an associated divergence exponent of the autocorrelation
length, $\nu=2$. 

\appendix

\section{Recoverable Energy as a Function of the Damage State}
\label{sec:derivation-of-the}

The argument here will be specific to  a fiber bundle model.  However,
as noted in the text, other weak damage models will also be controlled
by the same type of stored-energy function obtained here.

Define a fiber bundle as (initially) $N$ fibers stretched between
a free rigid plate and an elastic half-space.
The rigid plate has
a controlled displacement $l$ applied to it that stretches the fibers and
the elastic half-space.  As a fiber breaks at fixed $l$, the
force it held will be transmitted to the other fibers through the
elastic half-space.

A fiber at point ${\textbf{x}}$ is stretched a distance $\ell_{\textbf{x}}$.
Where that fiber is attached to the elastic halfspace, the surface
of the halfspace displaces an amount $u_{\textbf{x}}$. Thus, at those
places ${\textbf{x}}$ where fibers exist, we have
\begin{equation}
l=\ell_{\textbf{x}}+u_{\textbf{x}}.
\end{equation}
 The fiber at point ${\textbf{x}}$ exerts a force on the elastic
half-space that is
\begin{equation}
\frac{f_{\textbf{x}}}{A_{F}}=Y_{F}(1-\varphi_{\textbf{x}})\frac{\ell_{\textbf{x}}}{L_{F}}
\end{equation}
 where $A_{f}$ is the x-sectional area of the fiber (assumed to be
independent of the extension), $L_{F}$ is the inital length of each
fiber, and $Y_{F}$ is the Young's modulus of each fiber. The local
order parameter $\varphi_{\textbf{x}}$ is 0 if the fiber is intact
and 1 if broken.

The Green function for point forces acting on the surface of
an elastic halfspace \cite{Landau86} yields
\begin{eqnarray}
u_{\textbf{x}}=u_{3}({\textbf{x}}) & =
& \frac{1-\sigma_{s}^{2}}{\pi Y_{s}}
\sum_{{\textbf{y}}\neq{\textbf{x}}}\frac{f_{\textbf{y}}}{|{\textbf{y}}-{\textbf{x}}|}\\
 & = & \frac{1-\sigma_{s}^{2}}{\pi}\frac{Y_{F}}{Y_{s}}\frac{A_{F}}{L_{F}}
\sum_{{\textbf{y}}\neq{\textbf{x}}}
\frac{(1-\varphi_{\textbf{y}})}{|{\textbf{y}}-{\textbf{x}}|}\ell_{\textbf{y}}
\label{ux2}
\end{eqnarray}
 where $Y_{s}$ is the Young's modulus and $\sigma_{s}$ the Poisson's
ratio of the elastic solid. In general, the displacement at a point
${\textbf{x}}=(x_{1},x_{2},x_{3})$ within the elastic solid (where
$x_{3}=0$ defines the surface) due to a point force acting at a point
${\textbf{y}}=(y_{1},y_{2},0)$ on the surface
{[}i.e., ${\textbf{f}}({\textbf{x}})=f_{\textbf{y}}
\delta({\textbf{x}}-{\textbf{y}})\hat{\textbf{3}}${]}
is given by \begin{eqnarray}
\mbox{\hskip-5mm} u_{1} & \!\!\!=\!\!\! &
\frac{1+\sigma_{s}}{2\pi Y_{s}}
\left[\frac{(x_{1}-y_{1})x_{3}}{r^{3}}-
\frac{(1-2\sigma_{s})(x_{1}-y_{1})}{r(r+x_{3})}\right]f_{\textbf{y}} \label{u1}\\
\mbox{\hskip-5mm} u_{2}  & \!\!\!=\!\!\! & 
\frac{1+\sigma_{s}}{2\pi Y_{s}}
\left[\frac{(x_{2}-y_{2})x_{3}}{r^{3}}-
\frac{(1-2\sigma_{s})(x_{2}-y_{2})}{r(r+x_{3})}\right]f_{\textbf{y}}\label{u2}\\
\mbox{\hskip-5mm} u_{3}  & \!\!\!=\!\!\! & 
\frac{1+\sigma_{s}}{2\pi Y_{s}}
\left[\frac{x_{3}^{2}}{r^{3}}+
\frac{2(1-\sigma_{s})}{r}\right]f_{\textbf{y}}\label{u3}
\end{eqnarray}
 where
\begin{equation}
r=\left[(x_{1}-y_{1})^{2}+(x_{2}-y_{2})^{2}+x_{3}^{2}\right]^{1/2}.
\end{equation}
Putting $x_{3}=0$ in the expression for $u_{3}({\textbf{x}};{\textbf{y}})$
and then summing over all ${\textbf{y}}$ yields the  expression
for the displacement  $u_{\textbf{x}}$ of the surface.

We now define the dimensionless number
\begin{equation}
\varepsilon=\frac{(1-\sigma_{s}^{2})}{\pi}\frac{Y_{F}}{Y_{s}} \frac{A_F }{L_F L_P},
\end{equation}
where a length $L_P$  has been defined  as
\begin{equation}
\frac{1}{L_P} =  \sum_{\bf y \neq \bf x}  \frac{1}
{|{\bf y - \bf x}|};
\end{equation}
i.e., this sum is independent of which point ${\bf x}$ is considered.
Assuming either that the elastic halfspace is stiffer than the fibers, or  that
each fiber has  a length much great than its width,  or that fibers are spaced
far enough apart that  $L_p$ is large,  allows $\varepsilon$ to
be  considered a  small number.  Since the fiber bundle is assumed to be made of
a finite number $N$ of fibers, there is no divergence to $L_P$.

Using these definitions along with   $u_{\textbf{x}}=l-\ell_{\bf x}$
 and iterating Eq.\ (\ref{ux2}) once to get the leading order in $\varepsilon$
contribution gives
\begin{equation} \frac{\ell_{\bf x}}{l}
= 1 + \varepsilon\left(-1+  \sum_{\bf y \neq \bf x}
\frac{L_P}
{|{\bf y - \bf x}|}  \varphi_{\bf y}\right)  + O(\varepsilon^2).
\label{elloverU}
\end{equation}
The elastic strain energy reversibly stored in each surviving fiber is then
\begin{eqnarray}
\mbox{\hskip-6mm}  E_{\bf x} 
&=& \frac{1}{2} f_{\bf x} \ell_{\bf x}  =  \frac{1}{2}\frac{A_F Y_F}{L_F} l^2 
(1-\varphi_{\bf x}) \left(\frac{\ell_{\bf x}}{l}\right)^2 \\
\mbox{\hskip-6mm} &=& \frac{1}{2}\frac{A_F Y_F}{L_F}l^2 \left\{
1 - \varphi_{\bf x}  \phantom{\sum_{x}^{y}\left[\frac{L^2}{L^2}\right]} \right. \nonumber \\
&& \left. + 2  \varepsilon \!\left[-1 +\varphi_{\bf x}
+ \sum_{\bf y \neq \bf x}  \frac{L_P}
{|{\bf y - \bf x}|} \varphi_{\bf y} (1 - \varphi_{\bf x})
\right]\! \right\}.
\end{eqnarray}
where terms of $O(\varepsilon^2)$ have been dropped. 
Thus, upon summing over all the fibers we obtain the total energy $E_F$ stored in the
fibers as a function of the damage state
\begin{eqnarray}
&&\mbox{\hskip-9mm} E_F = \sum_{\bf x} E_{\bf x} = \frac{1}{2}\frac{A_F Y_F}{L_F} l^2 \nonumber \\
&&\mbox{\hskip-9mm} 
\times \!
\left[(1-2\varepsilon)N - (1-4 \varepsilon) \! \sum_{\bf x}\! \varphi_{\bf x}  - 
\varepsilon \!\!\!\sum_{{\bf x},{\bf y}\neq {\bf x}}\!\!\! J_{\bf x \bf y} 
 \varphi_{\bf x} \varphi_{\bf y}  \right]
\label{final}
\end{eqnarray}
where the coupling constant is defined 
\begin{equation}
J_{\bf x \bf y} = \frac{ 2 L_P}
{|{\bf y - \bf x}|}.
\end{equation}
This  form of the fiber energy is consistent with what was defined in
the text.

We now demonstrate that the
  energy  recoverably stored in the elastic solid makes no important
modification to $E_F$.  The strain energy in the solid is given by
 \begin{equation} E_s =
\frac{Y_s}{2(1+\sigma_s)} \int_{x_3 >0} d^3{\bf x} 
\sum_{\bf y \neq \bf x}  \left(e_{ik} e_{ik} + \frac{\sigma_s}{1-2\sigma_s} e_{ll}^2\right)
\label{es}
\end{equation}
where summation over the indices is assumed and where the strain tensor
is defined
\begin{equation}
e_{ik}({\textbf{x}};{\textbf{y}})
=\frac{1}{2}\left(\frac{\partial u_{i}({\textbf{x}};{\textbf{y}})}{\partial x_{k}}
+\frac{\partial u_{k}({\textbf{x}};{\textbf{y}})}{\partial x_{i}}\right).
\end{equation}
 The displacements are given by Eqs.\ (\ref{u1})--(\ref{u3}).

From  these equations, the strain at points ${\bf x}$ inside the elastic solid
takes the leading-order in $\varepsilon$  form
\begin{equation}
e_{ik} = \varepsilon l L_P \left[c_{ik} + \varepsilon \sum_{\bf y \neq \bf x}
\frac{f_{ik}({\bf x - \bf y})}{|{\bf x - \bf y}|^2} \varphi_{\bf y} \right]
\end{equation}
where the  constant tensor $c_{ik}$ has units of inverse-length squared  and
the average strain tensor throughout the elastic solid is
  $\varepsilon l L_P c_{ik}$.  The perturbation term due to broken fibers
volume integrates to zero.  The tensor $f_{ik}$ has no dependence on the
norm $|{\bf x - \bf y}|$; however, this fact is immaterial since $f_{ik}$
 plays no important role.

Upon forming the required products for the
integrand in Eq.\ (\ref{es}), and using the fact that terms linear in the
broken-fiber perturbations
integrate to zero, one obtains that the energy stored in the elastic solid is
\begin{equation}
E_s = \frac{1}{2}\frac{A_F Y_F}{L_F} l^2 \varepsilon  V L_p \left(c_{ik} c_{ik}
+ \frac{\sigma_s}{1-2\sigma_s} c_{ll}^2\right) \left[1 + O(\varepsilon^2)\right]
\end{equation}
where $V$ is the volume integrated over (assumed finite).
In otherwords, any energy stored in the elastic solid that is due to the
interaction between fibers,   is  $\varepsilon^2$ smaller than the leading-order
contribution which itself can be considered small.  The leading order contribution 
only depends on the average number of broken fibers and thus does not alter 
the analytical form of Eq.\ (\ref{final}).  
Thus, the energy stored in the elastic solid plays no essential role in the
damage model.

\bibliographystyle{apsrev}

\end{document}